\documentclass{article}
\usepackage{spconf,amsmath,graphicx}
\usepackage{listings}
\usepackage{bbm}
\usepackage{soul}
\usepackage{amsmath}
\usepackage{amsthm}
\usepackage{textcomp}
\usepackage{color, colortbl}
\usepackage{mathtools}
\usepackage{textcomp}
\usepackage{cancel} 
\usepackage{soul}
\usepackage{algorithm}
\usepackage{algpseudocode}

\usepackage{rh_defs_21}

\newcommand{\states}{\mathrm{states}}

\newcommand{\deformed}{\mathrm{deformed}}

\title{Non-rigid Motion Correction for MRI Reconstruction via Coarse-To-Fine Diffusion Models}
\name{Frederic Wang\textsuperscript{1} \qquad Jonathan I.\ Tamir\textsuperscript{2}}
\address{\textsuperscript{1}Department of Computing and Mathematical Sciences, Caltech \\
\textsuperscript{2}Chandra Family Department of Electrical and Computer Engineering, UT Austin}

\begin{document}
%
\maketitle
\begin{abstract}
Magnetic Resonance Imaging (MRI) is highly susceptible to motion artifacts due to the extended acquisition times required for $k$-space sampling. These artifacts can compromise diagnostic utility, particularly for dynamic imaging. We propose a novel alternating minimization framework that leverages a bespoke diffusion model to jointly reconstruct and correct non-rigid motion-corrupted $k$-space data. The diffusion model uses a coarse-to-fine denoising strategy to capture large overall motion and reconstruct the lower frequencies of the image first, providing a better inductive bias for motion estimation than that of standard diffusion models. We demonstrate the performance of our approach on both real-world cine cardiac MRI datasets and complex simulated rigid and non-rigid deformations, even when each motion state is undersampled by a factor of 64$\times$. Additionally, our method is agnostic to sampling patterns, anatomical variations, and MRI scanning protocols, as long as some low frequency components are sampled during each motion state.
\end{abstract}
\begin{keywords}
Magnetic resonance imaging, non-rigid, motion correction, diffusion, coarse-to-fine
\end{keywords}
\section{Introduction}
\label{sec:intro}

MRI produces images of high diagnostic versatility, but the scan time is very slow, often resulting in motion artifacts in the reconstructed image. Although compressed sensing methods \cite{lustig2007sparse,jalal2021robust} can accelerate scanning time, non-rigid motion may still occur, such as when scanning children or scanning anatomy with involuntary motion such as the heart, bowels, or lungs.

Image registration methods can correct motion by modeling it as a displacement field. Traditional methods involve fitting this field with B-splines or vector fields with regularization to ensure properties such as smoothness, diffeomorphism, and invertibility of the fitted displacement field \cite{chun2009simple, tustison2013explicit}. Deep learning approaches to image registration \cite{fu2020deep} can leverage data-driven anatomical priors to improve motion estimation, in particular for undersampled MRI data \cite{spieker2023deep}. Another common approach involves alternating between solving for the reconstructed image and estimating the motion \cite{aviles2021compressed}.

Diffusion models are a class of generative models that have achieved state-of-the-art performance for inverse problems such as inpainting, denoising, and MRI reconstruction \cite{chung2022diffusion, jalal2021robust}. However, leveraging diffusion models as a prior for non-rigid motion correction is still relatively unexplored.

\noindent\textbf{Our contributions:} 
\vspace{-2mm}
\begin{itemize}
\setlength{\parskip}{0pt} \setlength{\itemsep}{0pt plus 1pt}
  \item We develop a diffusion-based alternating minimization to jointly reconstruct a motion-corrected MRI image and corresponding non-rigid motion fields from a set of undersampled, motion-corrupted $k$-space. 
  \item We propose a novel coarse-to-fine diffusion model and show that it provides a better noise schedule for MRI reconstruction and registration by denoising the lower frequency $k$-space coefficients first. 
  \item Our method generalizes to arbitrary $k$-space sampling patterns with arbitrary motion without the need to re-train the model, as long as some low-frequency components of $k$-space are sampled at each motion state.
  \item While our method is developed to correct non-rigid motion, it can also correct for rigid motion comparable to the state-of-the-art.
\end{itemize}
\vspace{-2mm}
 We show the efficacy of our algorithm on complex retrospectively simulated motion as well as prospective motion corruption from cine cardiac MRI scans.

\subsection{Related Work} 
Several approaches have been proposed to leverage deep learning for non-rigid motion correction in MRI reconstruction \cite{spieker2023deep}. These methods typically involve simulating deformation fields for training, which naturally limits their use at inference time to motion that matches the simulation parameters. While this may be acceptable for regular, periodic motion such as respiratory and cardiac motion, it is not suitable for arbitrary motion a priori, as the learned model has been tuned to the particular motion patterns. Still, these approaches can be used given fully sampled training data \cite{pan2024motion}.

The use of diffusion models for MRI motion correction has been explored for rigid motion \cite{levac2023accelerated} by formulating the reconstruction as joint posterior sampling over the image and motion parameters. Diffusion models have also been used for more general blind inverse problems where the forward operator is unknown but belongs to a family of corruptions such as shift-invariant blurring \cite{chung2023parallel}, and this idea has also been applied specifically to MRI \cite{alkan2023variational}. Diffusion models on function spaces can also be used to enforce consistency between frames on video data \cite{daras2024warped}, which has potential applications in motion correction.

Diffusion models with non-isotropic noise have been studied for tasks such as image generation and posterior sampling. Coarse-to-fine blurring of diffusion models has been explored \cite{lee2022progressive} but only for prior sample generation. Diffusion bridges \cite{liu20232} also allow for non-isotropic noise for intermediate probability distributions. This can make sampling more efficient, particularly for solving inverse problems \cite{chung2024direct}.

\section{METHODS}
\label{sec:pagestyle}

\begin{figure}
    \centering
    \includegraphics[width = 0.95\columnwidth]{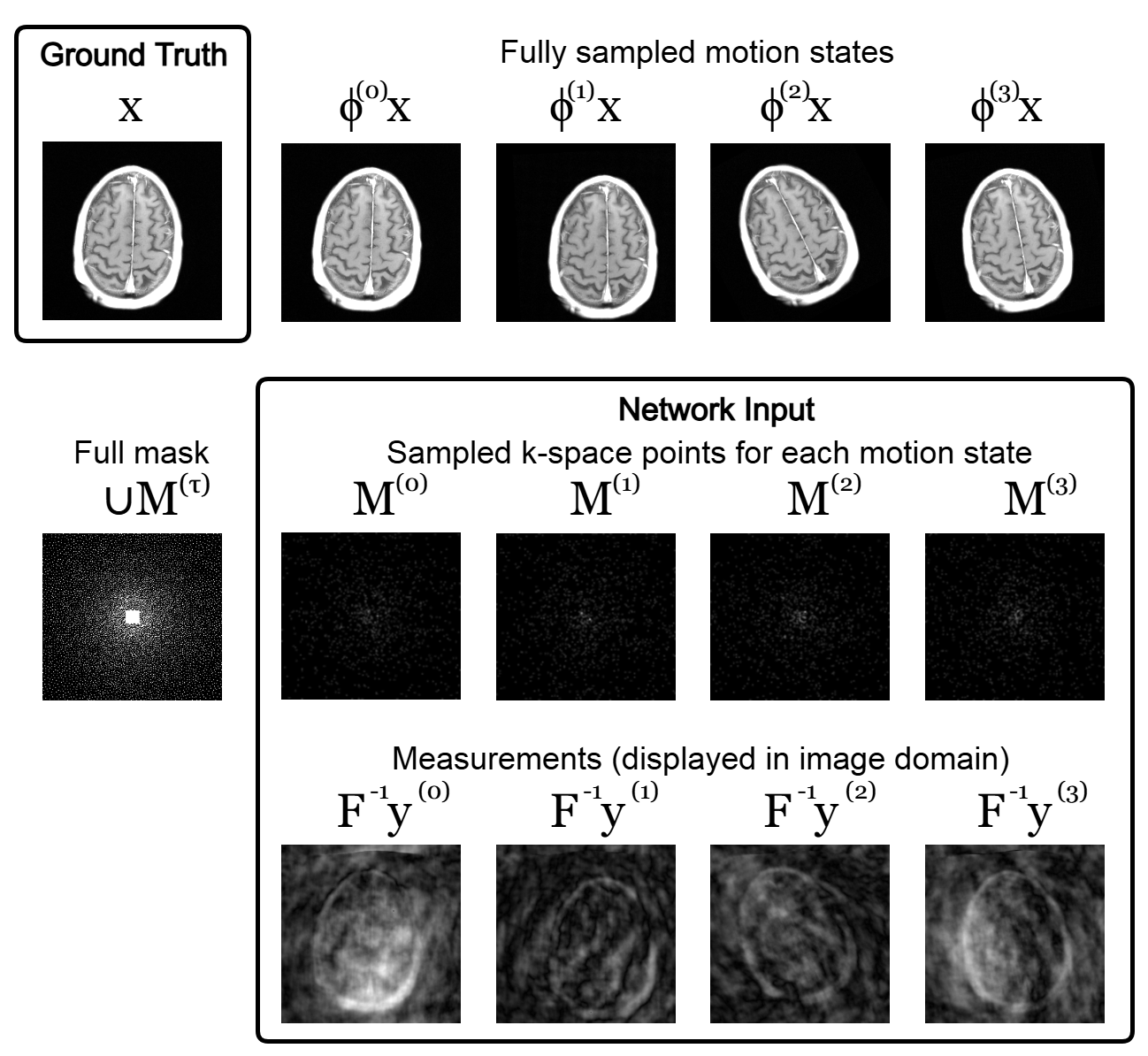}
    \caption{Visualization of the motion-corrupt forward model. Our goal is to recover the ground truth image $\vx$ (top left) given the undersampled, deformed measurements $\vy^{(\tau)}$ and undersampling masks $\mM^{(\tau)}$ (bottom right). Coils are omitted without loss of generality.}
    \label{fig:fig1_problem}
\end{figure}

\subsection{Problem Formulation}

Suppose we have motion-corrupted, multi-coil $k$-space from motion states $\{\vy_i^{(\tau)}\}$, where $\tau$ is a discrete variable that represents the motion state and $i$ represents the coil index. We assume $\tau=0$ is the motion-free state. The forward model for motion state $\tau \in \{0, \dots, n_{\states} - 1\}$ can be written as 
\begin{align}
    \vy_i^{(\tau)} &= \mM^{(\tau)} \mF \mS_i \Phi^{(\tau)} \vx + \varepsilon_i^{(\tau)}, \label{eq:forward}
\end{align}
where $\mM^{(\tau)}$ is the undersampling mask representing the $k$-space collected at motion state $\tau$; $\mF$ is the Fourier transform; $\mS_i$ is the sensitivity map of coil $i$; $\Phi^{(\tau)}$ is the displacement field for motion state $\tau$; and $\varepsilon_i$ is measurement noise. The sensitivity maps are a function of the MRI hardware and assumed constant over time. More succintly, 
\begin{align}
    \vy^{(\tau)} &= \mA^{(\tau)}\Phi^{(\tau)} \vx + \varepsilon^{(\tau)}, \label{eq:forward_simple}
\end{align}
where $\vy^{(\tau)}$ represents $k$-space across all coils and $\mA^{(\tau)}$ is the block-diagonal operator subsuming all coils from Eq. \ref{eq:forward}.
The goal is to reconstruct a fully sampled version of $\vx$ and the motion fields $\Phi^{(\tau)}$ given the deformed measurements $\vy^{(\tau)}$. See Fig. \ref{fig:fig1_problem} for a visual.


\subsection{Non-Isotropic Diffusion Models}
Given a clean image $\vx_0\sim p(\vx)$, the diffusion forward process of standard variance exploding (VE) \cite{song2020score} at time $t \in  [0, T]$ is 
\begin{align}
    \vx_t = \vx_0 + \mB_t \vz_t, \quad\vz_t \sim \mathcal{N}(0, \mI).
\end{align}
where $\vx_t$ is the intermediate corrupted image and $\mB_t$ is a linear operator that warps the noise. This process has an equivalent probability flow ODE
\begin{align}
    d\vx &= \left\{ \frac{d}{dt} \vx - \frac{d \mB_t}{dt} \mB_t \nabla_\vx \log p_t(\vx) \right\} dt,\label{eq:ODE}
\end{align}
where $p_t$ represents the distribution of $\vx_t$. A denoising network $D(\vx,t)$ trained with the standard score matching loss $\|D(\vx_t,t) - \vx_0\|_2^2$ can be used to approximate the score $\hat{s}(\vx_t, t) \approx \nabla_\vx \log p_t(\vx)$ via 
\begin{align}
    \hat{s}(\vx_t, t) = (\mB_t \mB_t^H)^{-1} (D(\vx;t) - \vx).
    \label{eq:score-estimation}
\end{align}
Once this network is trained, we can sample along the reverse process of this ODE to generate samples from $p(\vx)$. 

 We develop a coarse-to-fine (C2F) diffusion process by designing intermediate noise corruption masks $\mB_t$ to emphasize higher frequencies later and lower frequencies early on when the generated sample is still very noisy. This design accommodates multi-coil MRI applications, where the low frequencies of $k$-space are fully sampled to calibrate sensitivity maps (i.e. the auto-calibration region). Furthermore, this design also promotes better optimization in two ways: (1) the smoother optimization landscape of the low frequencies allows for smooth convergence towards the global minimum displacement field and (2) the low frequency areas are more robust to slight inaccuracies in the motion estimate during early stages of the algorithm.  

\begin{figure}
    \centering
    \includegraphics[width = 0.7\columnwidth]{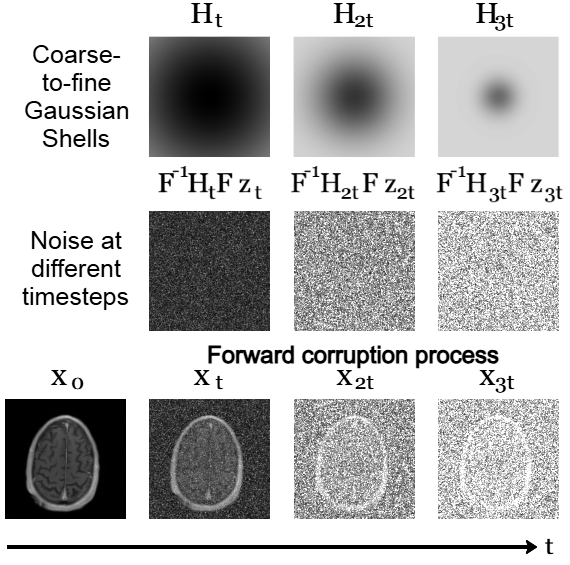}
    \caption{Forward diffusion process of Eq. \label{eq:forward_woo} where $t$ is some large timestep and $\mH_t$ is the frequency domain weighting for the noise at time $t$. As seen in the bottom row, more noise is added to the high frequency coefficients at the beginning.}
    \label{fig:fig2_forward_corruption}
\end{figure}

We define the intermediate corruption masks as $\mB_t = \mF^{-1}  \mH_t \mF$, where $\mH_t$ is an inverted Gaussian shell. Examples of these $\mH_t$ are shown in Fig. \ref{fig:fig2_forward_corruption}, top row. We propose Fourier weighting via Gaussian shells as they provide several advantages: (1) the windowing effect of the Gaussian decreases both ringing artifacts and aliasing in the noise, thus is more suitable for convolutional architectures such as the U-net backbone of the diffusion model; and (2) unlike the discrete masks chosen in other works, Gaussians are invertible, a necessary condition for unbiased sampling as derived from the score matching theory above.
This gives us the following diffusion process:
\begin{align}
    \vx_t &= \vx_0 + \mF^{-1}  \mH_t \mF \vz_t, \vz_t \sim \mathcal{N}(0, \mI).
    \label{eq:forward_woo2}
\end{align}
We can guide the diffusion model with a measurement $\vy$ by updating via $\nabla_{\vx} \log p_t(\vx_t | \vy)$ instead of $\log p_t(\vx_t)$. We can evaluate this via Bayes' rule using the approximation \cite{chung2022diffusion} $\log p_t(\vy | \vx_t) \approx \gamma \log p(\vy | \hat{\vx}_0)$, where $\hat{\vx}_0 = D(\vx_t, t)$ and $\gamma_t$ is a tunable parameter.
We summarize our sampling in Alg. \ref{alg:c2f}.
\begin{algorithm}[h!]
\caption{C2F Step($\vx_t$, $\vy$, $t$)}\label{alg:new_d}
\begin{algorithmic}
\State $\hat{\vx}_0 = D(\vx_t; t)$
\State $\hat{s}(\vx, t) = \mF^{-1} \mH_{t}^{-2} \mF (D(\vx;t) - \vx) + \gamma_t \nabla_{\vx} \log p(\vy | \hat{\vx}_0)$
\State $\vx_{t-1} \gets \vx_t + \mF^{-1}  (\mH_{t} - \mH_{t-1})  \mH_{t} \mF \hat{s}(\vx_t, t)$
\end{algorithmic}
\label{alg:c2f}
\end{algorithm}

\vspace{-4mm}
\subsection{Estimating the Displacement Field}

To estimate the displacement field between fully sampled image $\vx$ and deformed image $\vx_\deformed$,  we solve
\begin{align}
    \arg \min_{\Phi} \|\vx_{\deformed} - \Phi \vx\|_2^2 + R(\Phi),
    \label{eq:ideal_phi}
\end{align}
where $R(\Phi)$ is a regularization functional.
However, we are registering noisy and undersampled images; in particular, the displacement field from $\vx_t = \vx_0 + \vz_t$, $\vz_t \sim \mathcal{N}(0, F^{-1}  \mH_t^2 F)$, to the deformed $k$-space measurements $\vy$. There are two sources of errors compared to the ideal in Eq. (\ref{eq:ideal_phi}). 

First, we can only supervise with a low-rank version of the displacement field: namely $\mA\Phi$ rather than $\Phi$, but this error can be decreased via proper regularization. Our regularized estimation consists of two steps: (1) fitting a vector field via B-splines, enforcing smoothness and providing an approximate solution, and (2) pixel-wise gradient-based optimization while further regularizing with $\lambda \|\nabla \Phi\|_2^2$, where $\lambda$ is a fixed parameter.

The other error source is the diffusion corruption noise $\vz_t$. We can use our diffusion model to denoise $\vx_t$ by completing the reverse diffusion process, guiding the sampling with the motion-free measurement $\vy^{(0)}$. Crucially, we do not use the motion-corrupted measurements in this process, as that would require the current motion estimates, biasing the clean image and thus the next motion estimate closer to the current $\Phi$. The full motion algorithm can be seen in Alg. \ref{alg:motion}.

\begin{algorithm}[h!]
\caption{Update Motion($\bar{\vx}_t$, $\{ \vy^{(\tau)} \}$, $t$)}\label{alg:motion}
\begin{algorithmic}
\For {$s = t, \dots, 0$}
\State $\bar{\vx}_{s-1} \gets \text{C2F Step}(\bar{\vx}_s, \vy^{(0)}, s)$  
\EndFor
\For {$\tau = 0, \dots, n_{states}$}
\State $\Phi^{(\tau)} \gets \arg \min_{\Phi} \|\vy^{(\tau)} - \mA^{(\tau)} \Phi \bar{\vx}_{0}\|_2^2 + R(\Phi)$
\EndFor \\
\Return $\{ \Phi^{(\tau)} \}$
\end{algorithmic}
\end{algorithm}
\vspace{-4mm}
\subsection{Full Motion Correction Method}
Let $\mathcal{T}$ be a set of $N$ pre-selected reverse diffusion timesteps at which to update $\Phi^{(1)}, \dots, \Phi^{(n_\states - 1)}$. The full method can be seen in Alg. \ref{alg:c2f_full}. The motion update steps are nested within the reverse diffusion process, thus gradually improving the motion field estimate as the image gets cleaner.

\begin{algorithm}[h!]
\caption{Coarse-to-fine Diffusion Motion Correction}\label{alg:c2f_full}
\begin{algorithmic}
\State Input: $\{ \vy^{(\tau)} \}$, $\{ \mA^{(\tau)} \}$, $\mathcal{T} = \{t_1, \dots, t_N\}$
\State $\{ \Phi^{(\tau)} \} \gets 0$
\State $\vx_T \gets \mathcal{N}(0, F^{-1} \mH_T^2 F)$
\For {$t = T, \dots, 0$}
\If {$t \in \mathcal{T}$}
\State $\{ \Phi^{(\tau)} \} \gets$ Update Motion($\vx_t$, $\{ \vy^{(\tau)} \}$, $t$)
\EndIf
\State $\vx_{t-1} \gets \text{C2F Step}(\vx_t, \{ \vy^{(\tau)} \}, t)$
\EndFor \\
\Return $\vx_0$
\end{algorithmic}
\end{algorithm}
\vspace{-6mm}

\section{EXPERIMENTS}
\label{sec:typestyle}
\subsection{Simulated Data}
As a proof of principle of our method where motion-free ground-truth are available, we prepare raw multi-coil data from the fastMRI brain dataset \cite{knoll2020fastmri}, computing the coil sensitivity maps using ESPIRiT \cite{uecker2014espirit}. In order to train our network, we use 900 central slices from axial T1 post-contrast scans, cropped in $k$-space to 256x256. We then performed inference on 100 slices. We use this dataset to simulate both non-rigid and rigid motion. 

To simulate smooth, realistic non-rigid motion, we first generate random vector fields with Gaussian entries and then exponentially damp the energy in the Fourier domain. To simulate rigid motion, random rotation and translation values were generated for each motion state \cite{levac2023accelerated}. To simulate sampling masks, we first generate a 2D variable-density sampling mask. Afterwards, we partition this mask into disjoint subsets, representing the $k$-space collected at each motion state.

\subsection{Cine Cardiac MRI Data}
We used short axis cardiac cine data provided by the CMRxRecon2024 challenge \cite{wang2024cmrxrecon}. There were 180 subjects with fully sampled $k$-space; we used 150 subjects for training and 30 for inference. To generate training data, we used the first three temporal frames of the central slice of each subject, cropped in $k$-space to 162x162, for a total of 450 images. During inference, we used the provided Gaussian and uniform masks to undersample the data; these masks have an acceleration of R=24 each and there are a total of 12 motion states per scan. We emphasize that this is acquired data and thus there are no ground-truth motion fields.

\subsection{Implementation}
We used the EDM codebase \cite{karras2022elucidating} to train our diffusion model, keeping their noise schedule and preconditioning unchanged. The network uses separate channels for real and imaginary. We trained for 4M image steps on 8 V100 GPUs, and the training process for each network took about 48 hours. On the other hand, our motion correction method can range from 1 to 5 minutes on a single GPU for a single image at inference time, depending on the image resolution and the number of desired alternating minimization updates.

We compare our method to motion-informed posterior sampling (MI-PS) \cite{levac2023accelerated} using their public codebase\footnote{https://github.com/utcsilab/Motion\_Diffusion}. We used BART to run the baseline parallel image compressed sensing (PICS). \textbf{Ablation:} We also implemented a variation of our method but utilizing a standard diffusion model instead of the proposed C2F, which we call DPS-MC. For all methods, hyperparameters were tuned separately for a fair comparison. The posterior step hyperparameter $\gamma$ was tuned via grid search in the range [0.01, 100].

We parametrize the Gaussian shells via tunable parameters $a_t, \sigma_t$ for amplitude and standard deviation, respectively, for each timestep. We use the following formula for $\mH_t$:
\begin{align}
    \mH_t(x, y) = 1 - a_t \exp{\frac{-(x - H / 2)^2 - (y - W / 2)^2}{2 \sigma_t^2}}.
\end{align}
A schedule of $a_t = \max(\frac{1.1 t}{T}, 1)$ and $\sigma_t = 5 \exp (\frac{5 t}{T})$ performs well for generating both prior and posterior samples.

\section{Results}
\label{sec:majhead}

\begin{figure}
    \centering
    \includegraphics[width=0.88\columnwidth]{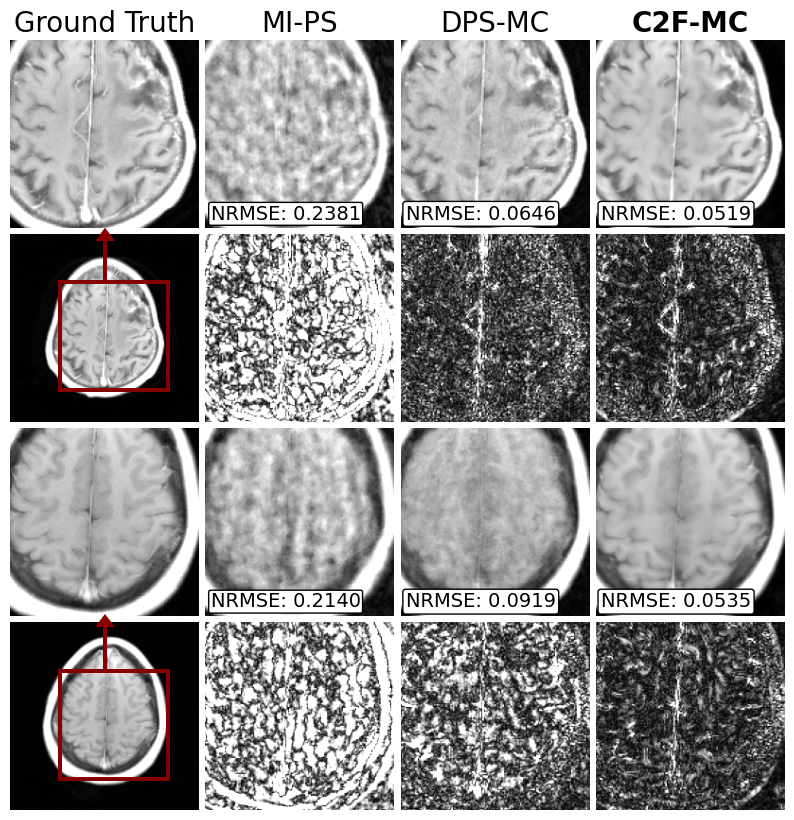}
    \caption{Non-rigid motion correction and reconstruction on multicoil data with 8 simulated motion states each undersampled by 64x. Error x10 is shown below each image. Our method (C2F-MC) outperforms the same method using standard diffusion (DPS-MC) and the state-of-the-art rigid motion correction (MI-PS).}
    \label{fig:multicoil-nonrigid}
\end{figure}

\subsection{Non-rigid simulated motion correction}
We test our method under the disjoint 2D sampling regime, designing $\mM$ to collect just 1/8 of $k$-space. However, we introduce 8 different motion states; thus, each $\mM^{(\tau)}$ has an acceleration of 64x, with a maximum displacement of at most 15 pixels. Results are shown in Fig. \ref{fig:multicoil-nonrigid}. We compare our results (C2F-MC) to the same method on a standard diffusion model (DPS-MC) and the state-of-the-art rigid motion correction (MI-PS). Both C2F-MC and DPS-MC used 10 alternating minimization iterations during inference, uniformly across the sampling process.

Our method outperforms all methods, demonstrating the effectiveness of coarse-to-fine denoising compared to standard diffusion methods. There is less hallucination and missing reconstruction details compared to the other methods. Furthermore, we perform better than MI-PS, as its rigid motion assumption does not generalize to non-rigid motion.

\noindent\textbf{Test Statistics:} Over 100 simulated non-rigid motion-corrupted images, the NRMSE (Normalized Root Mean Squared Error) statistic of C2F-MC was 0.058 $\pm$ 0.0032.

\begin{figure}
    \centering
    \includegraphics[width=1.0\columnwidth]{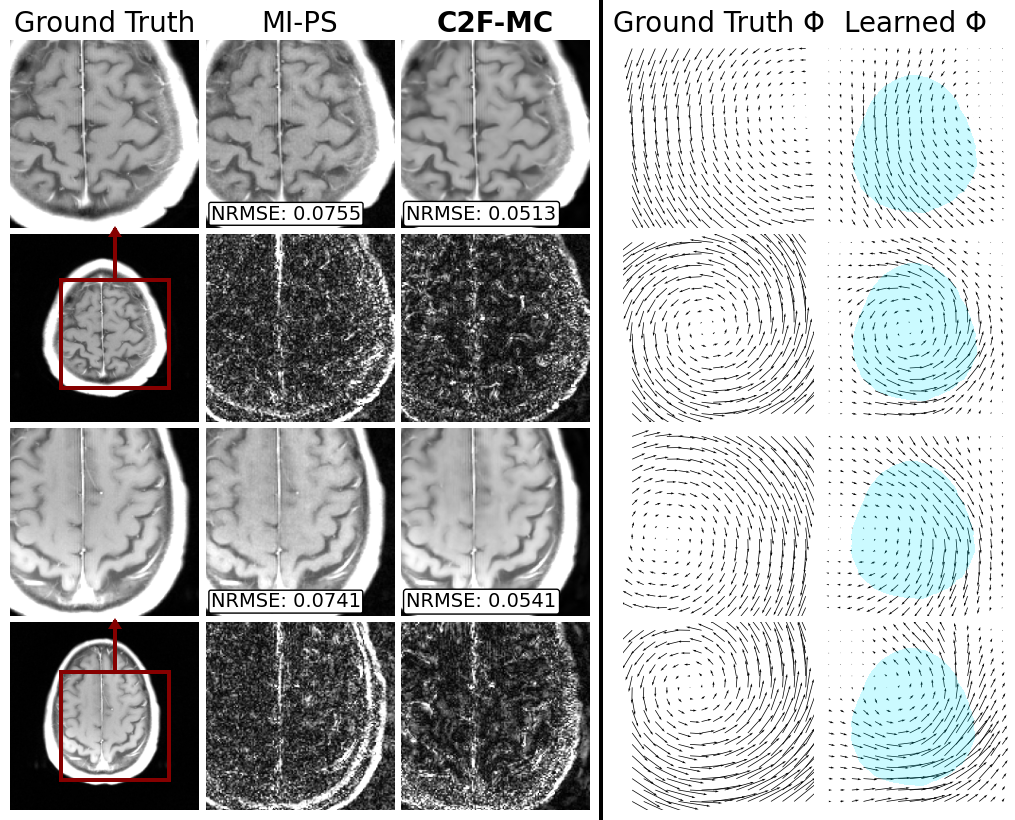}
    \caption{\textbf{Left:} Rigid motion correction and reconstruction on multicoil data with 8 motion states each undersampled by 64x. Error x10 is shown below each image. Our method (C2F-MC) slightly outperforms the state-of-the-art rigid motion correction (MI-PS). \textbf{Right:} Examples of the C2F-MC learned motion estimates and the corresponding ground truth motion. Our method is able to match rigid motion very well.}
    \label{fig:multicoil-rigid}
\end{figure}

\subsection{Rigid simulated motion correction}
We also tested our method under perfectly rigid motion. Once again, we assume a disjoint 2D sampling pattern with 8 motion states and an overall acceleration of R=8. The rigid motion parameters ($\theta$, $d_x$, $d_y$) were all chosen between [-5, 5] to simulate realistic motion. 

See Fig. \ref{fig:multicoil-rigid} for reconstruction results (left) and examples of how the learned motion estimates compare to the ground truth motion (right). Our method performs slightly better than the state-of-the-art (MI-PS) due to less ghosting artifacts near the edges. Furthermore, despite having substantially more parameters to learn (the whole motion field vs. three parameters), we closely match the ground truth rigid motion in the non-zero areas (blue highlight) of the image containing brain. 

\noindent\textbf{Test Statistics:} Over 100 rigid motion-corrupted images, the NRMSE of C2F-MC was 0.060 ± 0.0034, while the NRMSE of MI-PS was 0.082 ± 0.0028.

\begin{figure}
    \centering
    \includegraphics[width=0.84\columnwidth]{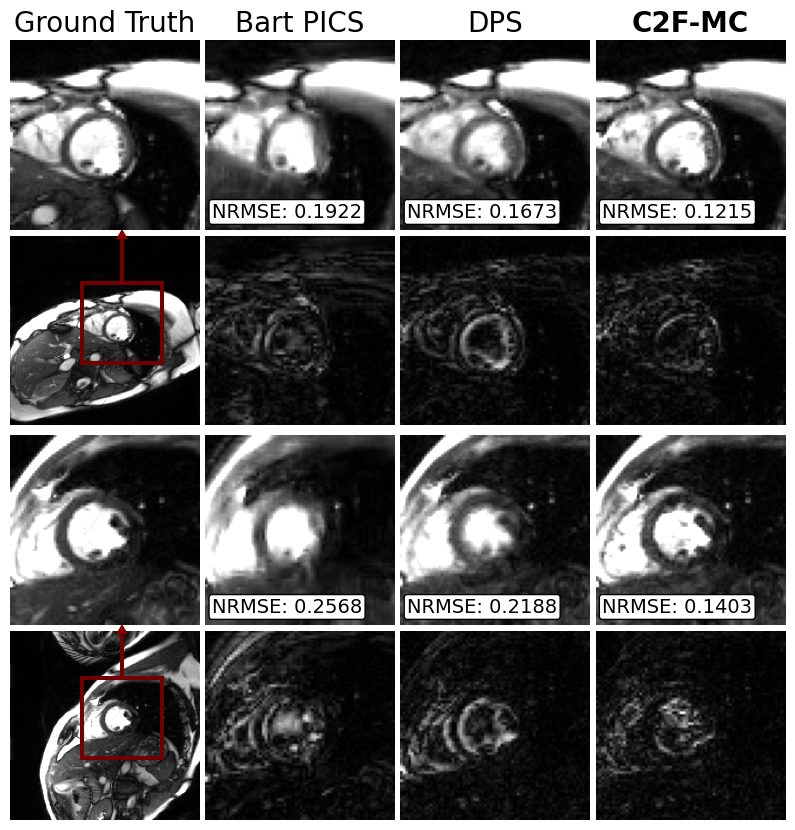}
    \caption{Reconstruction of cine Cardiac MRI. Error x10 is shown below each image. Our method (C2F-MC) outperforms the baseline (Bart PICS) as well as the reconstruction without motion correction (DPS).}
    \label{fig:cardiac-nonrigid}
\end{figure}

\subsection{Motion correction of cine cardiac MRI scans}
We also tested our method on real-world cine cardiac data which contains non-rigid motion near the aorta. The motion is localized within a small area in the image; thus, methods fitting global motion such as MI-PS are not appropriate. Each scan had 12 motion states each with acceleration of R=24.  We compared our results with PICS and standard diffusion posterior sampling without motion correction (DPS).
Visual results can be seen in Fig. \ref{fig:cardiac-nonrigid}. C2F both outperforms the baseline methods and is able to recover fine details within the heart.

\noindent\textbf{Test Statistics:} Over 30 cardiac scans, the NRMSE of C2F-MC was 0.1225 ± 0.0143.

\section{Conclusion}
\vspace{-3mm}
We introduced a method that leverages diffusion models to perform joint non-rigid motion correction and accelerated MRI reconstruction. We developed a coarse-to-fine diffusion model in order to use our inaccurate motion estimates to learn the low frequency areas of the $k$-space first. We also introduced a complementary alternating minimization algorithm to learn both the full-resolution image and the motion over time. The approach does not assume a specific motion model at training time and thus can be used in many applications where non-rigid motion is prevalent, such as cardiac imaging, while still having the ability to correct for global rigid motion, such as during brain scans.

\section{Acknowledgements}
\vspace{-3mm}
The authors thank Miki Lustig for fruitful discussions, and research support from NSF IFML 2019844, CCF-2239687, Google Research Scholars, and Chan Zuckerberg Initiative.

\bibliographystyle{IEEEbib}
\bibliography{references}

\end{document}